\documentclass[%
reprint,
superscriptaddress,
amsmath,amssymb,
aps,
pre,longbibliography
]{revtex4-2}

\usepackage{graphicx}
\usepackage{dcolumn}
\usepackage{bm}

\usepackage{mathtools}

\usepackage{color}

\usepackage[hyperfootnotes=false]{hyperref}

\usepackage{accents}
\usepackage{float}

\newtheorem{theorem}{Theorem}[section]
\newtheorem{remark}{Remark}

\begin{document}
	
	\title{A Class of Exclusion Processes Capable of Exhibiting Current Reversal}

	\author{Ngo Phuoc Nguyen Ngoc}
	\email{Coresponding author: ngopnguyenngoc@duytan.edu.vn}
	\affiliation{Institute of Research and Development, Duy Tan University, Da Nang, 550000, Vietnam.}
	\affiliation{Faculty of Natural Sciences, Duy Tan University, Da Nang 550000, Vietnam.}
	\author{Lam Thi Nhung}
		\email{nhunglt2@uef.edu.vn}
	\affiliation{Faculty of Information Technology, Ho Chi Minh City University of Economics and Finance, Ho Chi Minh, 700000, Vietnam}

	\begin{abstract}
			A century after Ising introduced the Ising measure to study equilibrium systems, its relevance has expanded well beyond equilibrium contexts, notably appearing in non-equilibrium frameworks such as the Katz--Lebowitz--Spohn (KLS) model. In this work, we investigate a class of generalized asymmetric simple exclusion processes (ASEP) for which the Ising measure serves as the stationary state. We show that the average stationary current in these models can display current reversal and other unconventional behaviors, offering new insights into transport phenomena in non-equilibrium systems. Moreover, although long-range interaction rates often give rise to long-range interactions in the potential function, our model provides a counterexample: even with long-range interactions in the dynamics, the resulting potential remains short-ranged. Finally, our framework encompasses several well-known models as special cases, including ASEP, the KLS model, the facilitated exclusion process, the cooperative exclusion process, and the assisted exchange model.
	\end{abstract}
	\maketitle
	
	
\section{Introduction}
Exactly solvable models are fundamental tools for exploring the physics of interacting many-body systems \cite{Baxter, Schutz2001}. Among them, the simple exclusion process (SEP) was first introduced in 1970 by Spitzer \cite{Spitzer1970} plays a central role in studying nonequilibrium dynamics in interacting particle systems \cite{Schutz2001, Spitzer1970, Katz,  Liggett1985, MacDonald1968,  Schutz2017,  Derrida1998, Blythe2007, Crampe2014, Golinelli2006}. This model describes particles performing asymmetric random walks on a lattice while obeying exclusion constraints. Despite its simplicity, the SEP captures a wide range of nonequilibrium phenomena, including traffic flow dynamics \cite{Schadschneider2000, Schadschneider2010, Chowdhury2000} and intracellular transport \cite{MacDonald1968, Belitsky2019-1, Belitsky2019-2}. Furthermore, it serves as a framework for studying key concepts in nonequilibrium statistical physics, such as the KPZ universality class \cite{Bertini1997, Sasamoto2010, Takeuchi2018} and boundary-induced phase transitions \cite{Schutz2001, Blythe2007}.

A distinctive feature of the SEP is its exact solvability, enabling the computation of physical quantities without approximations through methods such as the matrix product ansatz \cite{Derrida1998, Blythe2007} and the Bethe ansatz \cite{Schutz1993, Schutz1997, Liu2020, Tracy2008, Tracy2009}. However, the SEP remains an oversimplified representation of certain complex behaviors observed in nature, such as RNA polymerase pushing interactions in biological systems \cite{Belitsky2019-1, Belitsky2019-2, Dong2012, Dong2013, Epshtein2003-1, Epshtein2003-2} or current reversals in driven transport processes \cite{ Ngoc2025, Schulz2011, Chatterjee2018}. These limitations motivate the development of more generalized models incorporating additional interactions and dynamical rules to better describe real-world nonequilibrium phenomena.

The SEP models the movement of random particles on a lattice, subject to the constraint that no two particles can occupy the same site (exclusion principle). Particles can hop from one lattice site to a neighboring site, provided that the target site is unoccupied. Thus, the state of the process at each lattice site can be represented by a value in the set \( E = \{0,1\} \), where 0 denotes an unoccupied site and 1 represents an occupied site. The state space of the process on a lattice with \( L \) sites is \( \Omega_L = \{0,1\}^{L} \), where a configuration is denoted by \( \boldsymbol{\eta} = (\eta_{1}, \dots, \eta_{L}) \), with \( \eta_{i} \in \{0,1\} \). In one dimension, the transition rates are given by
\begin{equation}
	\accentset{\curvearrowright}{10} \to 01 \text{ with rate } r,\ \ 	\accentset{\curvearrowleft}{01} \to 10 \text{ with rate } \ell.
\end{equation}
The SEP has different variations depending on the hopping direction of the particles. If the particles are allowed to move in both directions with unequal hopping rates, $r$ and $\ell$, it is referred to as the \textit{Asymmetric Simple Exclusion Process} (ASEP). On the other hand, if the particles can only move in a single direction, either towards the right with $\ell = 0$ or towards the left with $r = 0$, it is known as the \textit{Totally Asymmetric Simple Exclusion Process} (TASEP). If both directions are allowed with equal hopping rates, i.e. $r = \ell$, the process is referred to as the \textit{Symmetric Simple Exclusion Process} (SSEP). Recall that the invariant measure of ASEP on the finite ring $\mathbb{T}_L: = \mathbb{Z}/L\mathbb{Z}$ is uniformly distributed. In the thermodynamic limit $L \to \infty$, the stationary particle current $j$ is the following 
\begin{align}\label{current_0}
	j = (r-\ell)\rho(1 - \rho),
\end{align}
where $\rho$ is the particle density \cite{Schutz2017, Grosskinsky}.

In SEP, particle interactions are limited to hardcore repulsion. However, as noted in \cite{Antal}, this constraint introduces unrealistic features in traffic flow modeling, such as the symmetric stationary current \eqref{current_0} and the absence of correlations in the steady-state particle distribution. Moreover, studies in biology and traffic models \cite{Belitsky2019-1, Epshtein2003-1, Epshtein2003-2} indicate that interactions among particles on the same lattice are inherently cooperative. Ignoring these interactions when multiple particles share a track is therefore impractical. The Katz-Lebowitz-Spohn (KLS) model \cite{Katz, Grosskinsky,  Schutz2016} addresses this by incorporating both hardcore repulsion and configuration-dependent jump rates, accounting for mutual particle interactions.

The KLS model  builds upon the ASEP by incorporating particle interactions. In the KLS model, the exclusion rule-preventing particles from occupying the same site-still applies, while interactions between particles are also considered. A generalized KLS model is investigated in \cite{Ngoc2025}, with its dynamics described as follows. A particle at site $i$ can move to site $i-1$ or $i+1$ with a rate that depends on the number of particles at sites $i-2$ and $i+1$ or $i-1$ and $i+2$, respectively, provided the target site is unoccupied. The dynamics can be visualized as follows:
\begin{widetext}
	\begin{equation}
	\begin{aligned}\label{GKLS_model}
		& 0\accentset{\curvearrowright}{10}0 \xrightarrow{r(1+\kappa)} 0010,\ 1\accentset{\curvearrowright}{10}0 \xrightarrow{r(1+\epsilon)} 1010,\ 0\accentset{\curvearrowright}{10}1 \xrightarrow{r(1-\epsilon)} 0011, \ 1\accentset{\curvearrowright}{10}1 \xrightarrow{r(1-\kappa)} 1011\\
		&0\accentset{\curvearrowleft}{01}0 \xrightarrow{\ell(1+\lambda)} 0100,\ 1\accentset{\curvearrowleft}{01}0 \xrightarrow{\ell(1-\epsilon)} 1100,\ 0\accentset{\curvearrowleft}{01}1 \xrightarrow{\ell(1+\epsilon)} 0101, \ 1\accentset{\curvearrowleft}{01}1 \xrightarrow{\ell(1-\lambda)} 1101
	\end{aligned}
\end{equation}
\end{widetext}
where $\epsilon, \kappa, \lambda \in (-1,1)$, and the text above the arrows indicates the jump rates. With the choice $\lambda = \kappa$, the model recovers the original KLS model. On the ring $\mathbb{T}_L$ with $L$ sites, the process admits the Ising measure with nearest-neighbor interaction energy as the invariant distribution, given by:
\begin{equation}\label{sm_1}
	\hat{\pi}(\boldsymbol{\eta}) = \dfrac{1}{Z}e^{-\beta J \sum_{i=1}^{L}\eta_i\eta_{i+1} + h\sum_{i=1}^{L}\eta_i}
\end{equation}
where $e^{\beta J} = \dfrac{1+\epsilon}{1-\epsilon}$ and $Z$ is the partition function. In equilibrium thermodynamics, the parameter $h$ represents the chemical potential that regulates the particle density, while the non-negative real parameter $\beta$ is proportional to the inverse of the positive temperature $T$, i.e., $\beta = 1/(k_B T)$, where $k_B$ is the Boltzmann constant, $J$ is the interaction strength between neighboring particles. A repulsive interaction corresponds to $J >0$, a non-interacting interaction takes place when $J=0$, and an attractive interaction is present when $J<0$.

An intriguing feature of this model is its ability to exhibit stationary current reversal, meaning the current can change sign as a function of density. As noted in \cite{Ngoc2025}, this phenomenon is not observed in the KLS model. However, it has been previously reported in the assisted exchange model \cite{Chatterjee2018}. The dynamics of the model are as follows:
\begin{equation}\label{aem}
	0\accentset{\curvearrowright}{10} \xrightarrow{p} 001, 0\accentset{\curvearrowleft}{01} \xrightarrow{q} 010,
	1\accentset{\curvearrowleft}{01} \xrightarrow{\gamma^{-1}q} 110, 1\accentset{\curvearrowright}{10} \xrightarrow{\gamma p} 101.
\end{equation}
This model serves as a particular instance of the framework outlined in \cite{Ngoc2025}, though it does not belong to the KLS model category.

The remainder of the paper is structured as follows. In the next section, we introduce the model, present its invariant measures, and highlight how it generalizes several well-studied models. In Section \ref{statiaonry_current}, we examine the stationary average current to illustrate key characteristics of the model. Notably, we demonstrate that the model can exhibit current reversal and display behaviors distinct from several well-known models. Finally, the last section provides concluding remarks. \\

\section{Models}
\subsection{Dynamics and invariant measure}\label{dynamics}
We consider a exclusion process on the finite ring $\mathbb{T}_L = \mathbb{Z}/L\mathbb{Z}$  with $L$ sites and $N$ particles. Before starting, we introduce some notation. Let $\textbf{p} = (p_1,p_2,...,p_N)$ represent the positions of the particles in a given configuration $\boldsymbol{\eta}$, where $p_i$ denotes the position of the $i$th particle on the lattice. Alternatively, the configuration $\boldsymbol{\eta}$ can be described by the headway vector $\textbf{h} = (h_1,...,h_N)$, where $h_i$ represents the number of empty sites between the $i$th and $(i+1)$th particles, i.e., $h_i = p_{i+1}-p_{i}-1$ mod $N$. 

Suppose that the jump rates of the $i$th particle to the right and left are denoted by $r_i(\boldsymbol{\eta})$ and $\ell_i(\boldsymbol{\eta})$, respectively. We assume that these rates depend on the positions of the two nearest neighbors of the particle. Specifically, the jump rates of particle 
$i$ are given by $r_i(\boldsymbol{\eta}) = r(h_{i-1}, h_i)$ and $\ell_i(\boldsymbol{\eta}) = \ell(h_{i-1}, h_i)$. Note that $r(0, k) = \ell(0, k) = 0$ due to the exclusion principle. We are now in a position to state the main theorem of the present paper. For the proof, see Appendix~\ref{proofs}.

\begin{widetext}
	\begin{theorem}\label{Th1}
	Consider a type exclusion process in which the jump rates of particles satisfy the following recursive relations:
	\begin{equation}\label{cond_jr_ep}
		\begin{cases}
			r(1,n) =  xr_{n+1},\ \ell(1,n) =  x\ell_{n+1}, \ \text{ for }n= 1,...,N-1,\\
			r(k,n) =  (xr_{n+k} +r_{n+k-1} +...+r_{n+1})-(r_{k-1}+...+r_1),\ \text{ for } k\geq 2, n\geq 1,\\
			\ell(k,n) =  (x\ell_{n+k} +\ell_{n+k-1} +...+\ell_{n+1})-(\ell_{k-1}+...+\ell_1),\ \text{ for } k\geq 2, n\geq 1.
		\end{cases}
	\end{equation}
	Here, the quantities $r_k := r(k,0), \ell_k := \ell(k,0)$ are free parameters. Under these conditions, the invariant measure of the process takes an Ising-like form:
	\begin{equation}\label{sm_Ising}
		\hat{\pi}(\boldsymbol{\eta}) = \dfrac{1}{Z}\pi(\boldsymbol{\eta}).
	\end{equation}
	where $\pi(\boldsymbol{\eta}) = x^{\sum_{i=1}^{L}\eta_i\eta_{i+1}}$ is the Boltzmann weight with nearest-neighbor interactions.

\end{theorem}
\end{widetext}

\begin{remark}
	An interesting interpretation of the parameter \( x \) emerges when we set  
	\begin{equation}
		x = e^{-\beta J},
	\end{equation}
	where \( \beta \) and \( J \) have the same meanings as in \eqref{sm_1}. Under this identification, the measure \eqref{sm_Ising} takes the same form as \eqref{sm_1}. Accordingly, a repulsive interaction corresponds to \( x < 1 \), the non-interacting case occurs when \( x = 1 \), and an attractive interaction is present when \( x > 1 \).
\end{remark}
\begin{remark}
	Another important observation is that, while the interaction range assumed in the jump rates \eqref{cond_jr_ep} is long, the interaction range in the Boltzmann weight $\pi(\boldsymbol{\eta})$ in \eqref{sm_Ising} remains short, involving only nearest-neighbor terms. This stands in contrast to the Headway Asymmetric Exclusion Process studied in \cite{Belitsky2024}, where long-range interactions in the jump rates lead to long-range interactions in the Boltzmann weight.
	
\end{remark}

In the following, we consider three particular cases of the model introduced in Theorem \ref{Th1}, with one, two, and three parameters corresponding to Models 1, 2, and 3, respectively.\\

\noindent\underline{\textbf{Model 1:}} Consider the case $r_1=r_2=...=r_{L-N}$ and $\ell_1=\ell_2=...=\ell_{L-N}$. The dynamics of the model can be visualized as follows.
\begin{equation}\label{dynamics_model1}
	1\accentset{\curvearrowright}{10} \xrightarrow{r_1} 101, 0\accentset{\curvearrowright}{10} \xrightarrow{xr_1} 001,
	\accentset{\curvearrowleft}{01}1 \xrightarrow{\ell_1} 101, \accentset{\curvearrowleft}{01}0 \xrightarrow{x\ell_1} 100.
\end{equation}

This model is known as cooperative exclusion process (as in \cite{Gabel2010}), where the particles exhibit totally asymmetric (i.e., unidirectional) movement, whereas our model allows particles to move in both directions. In the noninteracting case $x=1$ it becomes ASEP. In the limiting case, $x \to 0$, this model is a generalization of facilitated exclusion process \cite{Gabel2010, Basu2009,  Baik2016, Erignoux2024,Ayyer2023} with the following dynamics
\begin{equation}
	1\accentset{\curvearrowright}{10} \xrightarrow{r_1} 101, \quad
	\accentset{\curvearrowleft}{01}1 \xrightarrow{\ell_1} 101.
\end{equation}
In this limiting regime, the movement of a particle is strictly conditioned on the presence of another particle nearby, effectively enforcing a facilitation mechanism. Notably, the transition rules involving $x\ell$ and $xr$ vanish entirely, thereby preventing isolated particles from hopping. This behavior leads to a significantly constrained dynamics where movement is only possible in the presence of facilitating particles.\\

\noindent\underline{\textbf{Model 2:}} Consider the case  $r_2=r_3=....=r_{L-N}$ and $ \ell_2=\ell_3=....=\ell_{L-N}$ and $r_1, \ell_1, r_2, \ell_2$ are free parameters. The dynamics of the model can be visualized as follows.
\begin{widetext}
	
\begin{equation}\label{dynamics_model2}
	\begin{tabular}{c c c c}
		$ 1\accentset{\curvearrowright}{10}1 \xrightarrow{r_1} 1011$  & $1\accentset{\curvearrowright}{10}0 \xrightarrow{r_2} 1010$ &
		$0\accentset{\curvearrowright}{10}1 \xrightarrow{xr_2} 0011$ & $0\accentset{\curvearrowright}{10}0 \xrightarrow{(x+1)r_2 - r_1} 0010$ \\
		$ 1\accentset{\curvearrowleft}{01}1 \xrightarrow{\ell_1} 1011$  & $0\accentset{\curvearrowleft}{01}1 \xrightarrow{\ell_2} 1010$ &
		$1\accentset{\curvearrowleft}{01}0 \xrightarrow{x\ell_2} 0011$ & $0\accentset{\curvearrowleft}{01}0 \xrightarrow{(x+1)\ell_2 - \ell_1} 0010$ 
	\end{tabular}
\end{equation}

\end{widetext}
This model extends the generalized KLS model \eqref{GKLS_model} with a choice $x = \dfrac{1-\epsilon}{1+\epsilon}$, thereby providing a broader framework that encompasses several well-known models--including the assisted exchange model \eqref{aem}, in which current reversal also occurs.\\

\noindent\underline{\textbf{Model 3:}} Consider the case  $r_3=....=r_{L-N}$ and $ \ell_3=....=\ell_{L-N}$ and $r_1, r_2, \ell_1, \ell_2, r_3, \ell_3$ are free parameters. The dynamics of the model can be visualized as follows.
\begin{widetext}
\begin{equation}\label{model3}
	\begin{tabular}{|c| c| c| c|}
		\hline
		Rightward Transition & Rate & Leftward Transition & Rate\\
		\hline
		$1\accentset{\curvearrowright}{10}1 \to 1011$ & $r_1$ & 
		$1\accentset{\curvearrowleft}{01}1 \to 1101$ & $\ell_1$ \\
		$1\accentset{\curvearrowright}{10}01 \to 10101$ & $r_2$ & $10\accentset{\curvearrowleft}{01}1 \to 10101$ & $\ell_2$ \\
		$1\accentset{\curvearrowright}{10}00 \to 10100$ & $r_3$ &
		$00\accentset{\curvearrowleft}{01}1 \to 00101$ & $\ell_3$ \\
		$10\accentset{\curvearrowright}{10}1 \to 10011$ & $xr_2$ & $1\accentset{\curvearrowleft}{01}01 \to 11001$ & $x\ell_2$\\
		$ 00\accentset{\curvearrowright}{10}1 \to 00011$ &$xr_3$ & $1\accentset{\curvearrowleft}{01}00 \to 11000$ & $x\ell_3$\\
		$00\accentset{\curvearrowright}{10}01 \to 000101$ & $(x+1)r_3-r_1$ & $10\accentset{\curvearrowleft}{01}00 \to 101000$ & $(x+1)\ell_3-\ell_1$\\
		$10\accentset{\curvearrowright}{10}00 \to 100100$ & $(x+1)r_3-r_1$ & $00\accentset{\curvearrowleft}{01}01 \to 001001$ & $(x+1)\ell_3-\ell_1$ \\
		$ 10\accentset{\curvearrowright}{10}01 \to 100101$ & $xr_3+r_2-r_1$ & $10\accentset{\curvearrowleft}{01}01 \to 101001$
		& $x\ell_3+\ell_2-\ell_1$\\
		$00\accentset{\curvearrowright}{10}00 \to 000100$ & $(x+2)r_3-r_2-r_1$ & $00\accentset{\curvearrowleft}{01}00 \to 001000$ &$(x+2)\ell_3-\ell_2-\ell_1$\\
		\hline
	\end{tabular}
\end{equation}
\end{widetext}
This model introduces an additional level of complexity and has not been previously discussed in the literature. Clearly, Model 3 generalizes Model 2, and Model 2 generalizes Model 1. The following remark shows that a broader class of exclusion processes, beyond the one in Theorem \ref{Th1}, admits the Ising-like measure \eqref{sm_Ising} as its invariant measure.

\begin{remark}\label{Th3}
	Suppose that $r(k,n)$ and $\ell(k,n)$, for $k\geq 1$ and $n \geq 0$, are given as in Theorem \ref{Th1}. Consider a class of exclusion processes with rightward and leftward jumps, given as follows:
	\begin{equation}\label{cond_jr_ep1}
		\begin{cases}
			u_r(k,n) = \lambda[r(k,n)] + \gamma x r_k r_{n+1}, \\
			u_{\ell}(k,n) = \beta[\ell(k,n)] + \gamma x r_k r_{n+1},
		\end{cases}
	\end{equation}
	for $ k\geq 1, n\geq 0 $ and where $\lambda, \beta$ and $\gamma$ are non-negative parameters. Then, the invariant measure of the process follows an Ising-like distribution as in \eqref{sm_Ising}. Notice that, with the choice $\lambda = \beta = 1$ and $\gamma = 0$, this class of exclusion processes reduces to the one considered in Theorem \ref{Th1}. This implies that the conditions on the jump rates in Theorem \ref{Th1} are not necessary for the exclusion process to admit the Ising-like measure as its invariant distribution.
\end{remark}

The model in Theorem \ref{Th1} is generally irreversible, except in the specific case described below. Physically, the kinetic parameters \( r_i \) and \( \ell_i \) represent nonreciprocal interactions that violate Newton’s third law--phenomena commonly observed in complex systems ranging from intracellular transport to collective behavior in living organisms \cite{Bowick2022}. Analyzing the stationary states of such systems remains a major challenge.

Identifying the conditions under which reversibility arises is essential for understanding these dynamics. Reversibility guarantees the existence of a stationary distribution satisfying detailed balance, which simplifies analysis and links the system to equilibrium statistical mechanics. The following theorem provides necessary and sufficient conditions for reversibility, emphasizing the role of kinetic parameters.

\begin{theorem}\label{Th2}
	Consider a process governed by the dynamics given in \eqref{cond_jr_ep} of Theorem \ref{Th1}. This process is reversible with respect to the measure \eqref{sm_Ising} if and only if \( r_1 = \ell_1,\, r_2 = \ell_2,\, \ldots,\, r_{L-N} = \ell_{L-N} \).
\end{theorem}

For the proof of the above theorem, see Appendix~\ref{proofs}. According to the theorem, if there exists an index \( i \) such that \( r_i \neq \ell_i \), then the sources are unbalanced, and in the steady state, there will be a net flux of particles through the system.

\section{Stationary average current and discussion}\label{statiaonry_current}
\subsection{Average current}
The primary quantities that characterize the nonequilibrium nature of the model are the average stationary velocity and the particle current.  The average velocity \( \nu \)  is defined as the expected value of the difference between the rightward and leftward jump rates, as established in \cite{Belitsky2019-1, Bahadorana2015}:
\begin{equation}
	\nu = \langle r_i(\boldsymbol{\eta}) -\ell_i(\boldsymbol{\eta})\rangle_{\hat{\pi}},
\end{equation}
where \( \langle\cdot \rangle_{\hat{\pi}} \) denotes the expectation with respect to the stationary measure \( \hat{\pi} \) given by equation \eqref{sm_Ising}. The stationary flux \( j \) and the stationary velocity \( \nu \) are related to the particle density \( \rho \) by the equation:
\begin{equation}
	j = \rho \nu.
\end{equation}

\begin{figure}[h]
	\centering
	\includegraphics[width=8cm]{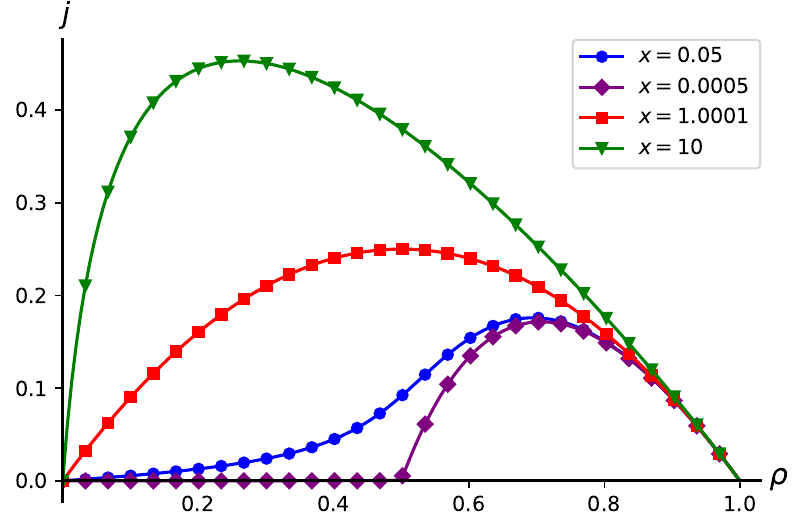}
	\caption{Stationary current $j$ of Model~1 versus particle density $\rho$ in the repulsive ($x = 0.05$ and $0.0005$), non-interacting ($x = 1.0001$), and attractive ($x = 10$) regimes. The jump rates are $r_1 = 2$ and $\ell_1 = 1$.
	}
	\label{current_mod1}
\end{figure}

\begin{figure}[h]
	\centering
	\includegraphics[width=8cm]{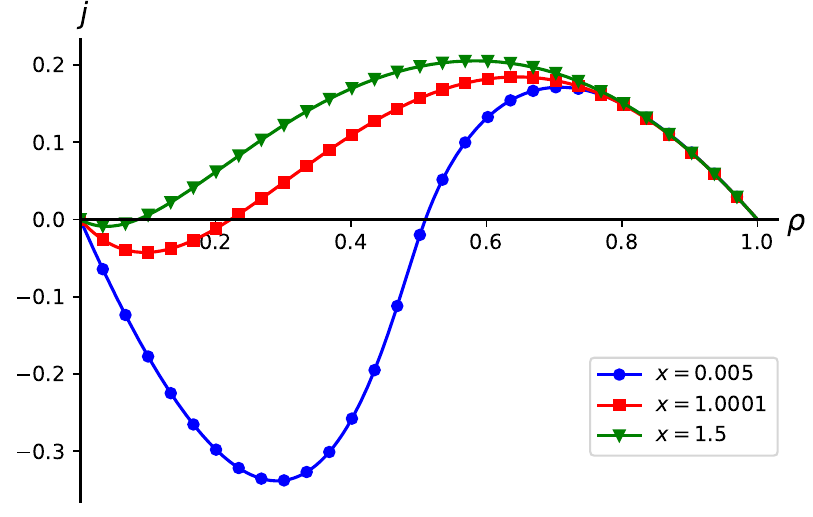}
	\caption{Stationary current $j$ of Model~2 versus particle density $\rho$ in the repulsive ($x = 0.005$), non-interacting ($x = 1.0001$), and attractive ($x = 10$) regimes. The jump rates are $r_1 = 4$, $r_2 = 2$, and $\ell_1 = \ell_2 = 1$.
	}
	\label{current_mod2}
\end{figure}

\begin{figure}[h]
	\centering
	\includegraphics[width=8cm]{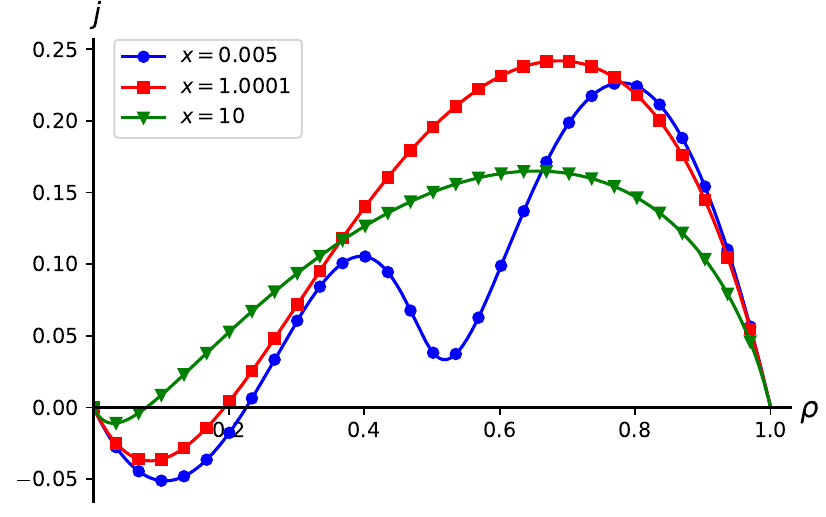}
	\caption{Stationary current $j$ of Model~3 versus particle density $\rho$ in the repulsive ($x = 0.005$), non-interacting ($x = 1.0001$), and attractive ($x = 10$) regimes. The jump rates are $r_1 = 1$, $r_2 = 2$, $r_3 = 4$, and $\ell_1 = 2$, $\ell_2 = 0.005$, $\ell_3 = 4$.
	}
	\label{current1}
\end{figure}

\begin{figure}[h]
	\centering
	\includegraphics[width=8cm]{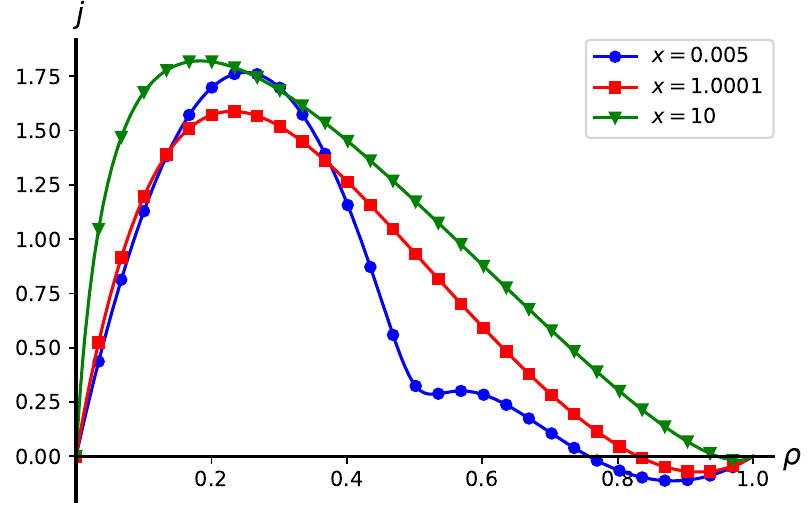}
	\caption{Stationary current $j$ of Model~3 versus particle density $\rho$ in the repulsive ($x = 0.005$), non-interacting ($x = 1.0001$), and attractive ($x = 10$) regimes. The jump rates are $r_1 = 0.0001$, $r_2 = 0.1$, $r_3 = 5$, and $\ell_1 = 4$, $\ell_2 = 2$, $\ell_3 = 1$.
	}
	\label{current2}
\end{figure}

\begin{figure}[h]
	\centering
	\includegraphics[width=9cm]{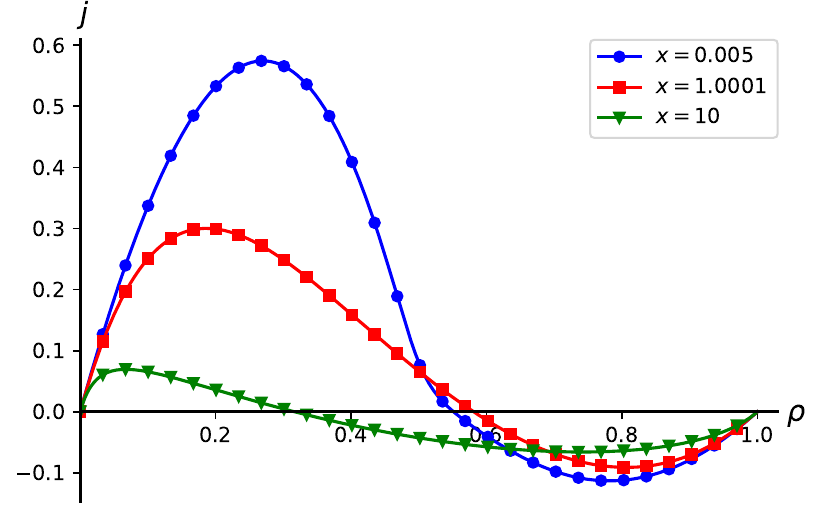}
	\caption{Stationary current $j$ of Model~3 versus particle density $\rho$ in the repulsive ($x = 0.005$), non-interacting ($x = 1.0001$), and attractive ($x = 10$) regimes. The jump rates are $r_1 = r_2 = r_3 = 1$, and $\ell_1 = 4$, $\ell_2 = 2$, $\ell_3 = 1$.
	}
	\label{current3}
\end{figure}

To compute the average current, it is convenient to use the grand-canonical ensemble \eqref{inv_meas_gc_2} together with the headway distribution \eqref{meanheadway} (see Appendix~\ref{average_current}). Specifically,
\begin{equation} j = \rho \langle r_i(\boldsymbol{\eta}) - \ell_i(\boldsymbol{\eta}) \rangle_{\tilde{\pi}_{gc}}, \end{equation}
where the expectation $\langle \cdot \rangle_{\tilde{\pi}_{gc}}$ is taken with respect to the grand-canonical measure \( \tilde{\pi}_{gc} \) defined in equation \eqref{inv_meas_gc_2}. The headway distribution \eqref{meanheadway} is then used to evaluate this expression, enabling the average particle current $j$ to be plotted; see Fig. \ref{current_mod1} for Model 1, Fig. \ref{current_mod2} for Model 2, and Figs. \ref{current1}--\ref{current3} for Model 3.


\subsection{Discussion}

The analysis of the stationary average current reveals several intriguing features that distinguish our model from classical counterparts. One of the most striking is the phenomenon of \textit{current reversal}, in which the stationary current changes sign as a function of particle density, as illustrated in Figs.~\ref{current_mod2}–\ref{current3}. Notably, this reversal occurs even in the non-interacting case, i.e., when the interaction strength is \( J = 0 \), ($x=1$). This suggests that the reversal originates not from the interactions encoded in the Boltzmann weight $\pi(\boldsymbol{\eta})$ in \eqref{sm_Ising}, but from the structure of the jump rates themselves.

Current reversal does not occur in Model 1 \eqref{dynamics_model1}, which includes many classical models as special cases. Specifically, the average current in this model can be computed using the headway distribution \eqref{meanheadway}:
\begin{equation} 
	j = (r_1 - \ell_1)\dfrac{xz}{(x + (1 - x)z)^2} 
\end{equation} 
where $z$ is defined in \eqref{value_z}. The sign of $j$ depends solely on the difference $r_1 - \ell_1$, confirming that current reversal is absent in Model 1 (see Fig. \ref{current_mod1}). In contrast, this phenomenon does arise in Model 2, as shown in Fig. \ref{current_mod2}.

Another interesting observation is that current reversal can occur at both low and high particle densities. See Figs. \ref{current_mod2} and \ref{current1} for the former case, and Figs. \ref{current2} and \ref{current3} for the latter. This behavior contrasts with the Assisted Exchange Model, where current reversal typically occurs around the point $\rho = 0.5$ (see Fig. 1 in \cite{Chatterjee2018}).

A fundamental difference between the ASEP and our model lies in the symmetry properties of the current profile. In ASEP, the current is symmetric with respect to the particle density $\rho = 0.5$, remaining invariant under the transformation $\rho \to 1 - \rho$, and typically exhibits a single maximum. In contrast, our model does not exhibit such symmetry; the current profile changes under the same transformation. Consequently, the current can display multiple extrema. For example, in Model 3 under repulsive interactions, the current may feature up to four extrema--two maxima and two minima (see Fig. \ref{current1}). This behavior surpasses that of the KLS model, which allows at most three extrema. The emergence of additional extrema in our model highlights a novel transport feature, reflecting a more complex interplay between particle interactions and the structure of jump rates. This complexity gives rise to multiple density regimes in which the current varies in a highly nontrivial manner.

These findings highlight the uniqueness of our model and its potential implications for understanding transport phenomena in driven systems. The ability to generate multiple extrema and current reversals opens new directions for exploring non-equilibrium steady states, particularly in systems with competing interaction mechanisms.

\section{Conclusions}
In this work, we introduced and analyzed a class of generalized ASEP models in which the Ising measure serves as the stationary distribution. Our results highlight the rich dynamical properties of these systems, particularly the emergence of current reversal and unconventional transport behavior. These findings offer deeper insights into non--equilibrium systems and the mechanisms governing their transport dynamics.

Furthermore, while the jump rates in our construction involve long-range dependencies, the resulting stationary (Boltzmann) weight can still be written in a short-range, nearest-neighbor form. This demonstrates that long-range kinetic interactions do not necessarily generate explicit long-range coupling terms in the invariant measure, thus providing a concrete counterexample to conventional expectations. This clarification refines the conventional understanding of the relation between the interaction range encoded in the jump dynamics and that emerging in the stationary Boltzmann weight.

Beyond their theoretical significance, our models encompass and generalize several well-known exclusion processes, including ASEP, the KLS model, the facilitated exclusion process, the cooperative exclusion process, and the assisted exchange model. This unifying framework underscores the versatility of our approach and points to potential applications in modeling transport phenomena across a range of physical and biological systems.

\section*{Appendices}\label{app}
\appendix
\renewcommand{\thesubsection}{\Alph{section}.\arabic{subsection}}

\section{Headway distribution }\label{app_hw}

Since the number of particles is fixed in our models, we employ the grand-canonical ensemble for analytical convenience:
\begin{equation}\label{inv_meas_gc2}
	\tilde{\pi}_{gc}(\boldsymbol{\eta}) = \dfrac{1}{Z_{L,N}}e^{-\frac{1}{k_B T} \left( J\sum_{i=1}^{L}\eta_i\eta_{i+1} + h\sum_{i=1}^{L}\eta_i\right)},
\end{equation}
instead of \eqref{sm_Ising}, where the parameter $h$ controls the average particle density.

Using translation invariance, a configuration $\boldsymbol{\eta} = (\eta_1,...,\eta_L)$ is represented by the headway vector $\textbf{h} = (h_1,...,h_N)$. We define  
\begin{equation}\label{newvar} 
	\theta_{i}^r := \delta_{h_i, r}, 
\end{equation}  
where $i$ is taken modulo $N$ and $\delta$ denotes the Kronecker delta, defined by  
\begin{equation}\label{kro_intro}
	\delta_{\alpha,\beta} =
	\begin{cases}
		1, & \text{if } \alpha = \beta,\\
		0, & \text{otherwise},
	\end{cases}
\end{equation}  
for $\alpha, \beta$ belonging to any set. The ensemble \eqref{inv_meas_gc2} then transforms into
\begin{equation}\label{inv_meas_gc_2}
	\tilde{\pi}_{gc}(\textbf{h}) =\dfrac{1}{\tilde{Z}}\prod_{i=i}^{N}x^{\theta_i^0}z^{h_i},
\end{equation} 
with partition function
\begin{equation} \tilde{Z} = \left(\frac{1+(1-x)z}{1-z}\right)^N. 
\end{equation} 
To ensure the well-definedness of the grand-canonical measure \eqref{inv_meas_gc2}, we identify the fugacity $z$ corresponding to density $\rho$. Since $\tilde{Z}$ must be finite, the fugacity must satisfy $0\leq z <1$. The mean headway is
\begin{equation} 
	\langle h_i\rangle_{\tilde{\pi}_{gc}} = \frac{z}{(1-z)(x+(1-x)z)}, 
\end{equation}
which equates to the lattice-based computation $\langle h_i\rangle_{\tilde{\pi}_{gc}} = \dfrac{1}{\rho} -1$, leading to
\begin{equation} \left(\frac{1}{x}-1\right)z^2 + z\left( \frac{1}{x}\frac{\rho}{1-\rho} - \frac{1}{x} +2\right) -1 =0. \end{equation} 
Solving for $z$:
\begin{equation}\label{value_z} z:=z(\rho,x)= 1 -\frac{1 - \sqrt{1 - 4\rho(1-\rho)(1-x) }}{2(1-\rho)(1-x)}. \end{equation} Since $\rho \in [0,1)$, the measure remains well-defined.

For the headway process, the probability distribution is
\begin{equation} 
	P_h(r) =\tilde{\pi}_{gc}(h_i = r) = \langle \theta_i^r\rangle_{\tilde{\pi}_{gc}}. 
\end{equation} Explicitly,
\begin{equation}\label{meanheadway} 
	P_h(r) = \begin{cases} \dfrac{(1-z)x}{x+(1-x)z}, & r = 0, \\ 
		\dfrac{1}{x}P_h(0)z^r, & r \geq 1. 
	\end{cases} 
\end{equation}

In the thermodynamic limit, it is straightforward to show that the headway random variables $\theta_{i-1}$ and $\theta_i$ become independent. Specifically, one has $\langle \theta_{i-1} \theta_i \rangle_{\tilde{\pi}_{gc}} = \langle \theta_{i-1} \rangle_{\tilde{\pi}_{gc}} \langle \theta_i \rangle_{\tilde{\pi}_{gc}}$ for $i = 1, \dots, N$. This observation facilitates the computation of the stationary average current of the model.

\section{Average current of Models 1,2,3}\label{average_current}
\noindent \underline{\textbf{Model 1:}} The jump rates \eqref{dynamics_model1} can be expressed in a general parameter form that is valuable in calculating the average particle current. Namely, the rates read off
\begin{equation}\label{rate_md_2}
	\begin{cases}
		r_i(\boldsymbol{\eta})  =  r_1^\star(1+d_r^{1\star}\theta_{i-1}^0 )(1-\theta_{i}^0) \\
		\ell_i(\boldsymbol{\eta}) =  \ell_1^\star(1+d_\ell^{\star1}\theta_{i}^0 )(1-\theta_{i-1}^0) 
	\end{cases}
\end{equation}
where $r_1^\star = xr_1$, $\ell_1^\star = x\ell_1$ and $d_r^{1\star} = d_\ell^{\star1} = \dfrac{1}{x}-1$. 
The superscript $\star$ designates the position of particle $i$, and the symbols 0 and 1 denote unoccupied and occupied sites, respectively. As for the parameter $d^{1\star}_r$, it affects the rate $r_{i}(\boldsymbol{\eta})$ only when particles $i-1$ and $i$ occupy neighboring sites. The factor $\theta_{i-1}^0$ encodes this adjacency, taking the value 1 only when the particles are adjacent.  The parameter $d^{\star1}_\ell$ is interpreted in a similar manner.

Thus the average current for this model is computed as follows
\begin{equation}\label{current_model1}
	\begin{aligned}
		j & =\  \langle r_i(\boldsymbol{\eta}) - \ell_i(\boldsymbol{\eta}) \rangle_{\tilde{\pi}_{gc}}\\
		&=\  r_1^\star(1+d_r^{1\star}\langle \theta_{i-1}^0 \rangle_{\tilde{\pi}_{gc}} )(1-\langle \theta_{i}^0 \rangle_{\tilde{\pi}_{gc}}) \\
		&\quad -  \ell_1^\star(1+d_\ell^{\star1}\langle \theta_{i}^0 \rangle_{\tilde{\pi}_{gc}} )(1-\langle \theta_{i-1}^0 \rangle_{\tilde{\pi}_{gc}}) 
	\end{aligned}
\end{equation}
Thus, using headway distribution \eqref{meanheadway} to get explicit formula of average current $j$.\\

\noindent \underline{\textbf{Model 2:}} One can rewrite the rates \eqref{dynamics_model2} in a unified parameter form as follows
\begin{equation}\label{rate_md_3}
	\begin{cases}
		r_i(\boldsymbol{\eta}) =\  r_2^\star(1+e_r^{1\star}\theta_{i-1}^0 + e_r^{\star 01}\theta_{i}^1 )(1-\theta_{i}^0),\\
		\ell_i(\boldsymbol{\eta}) =\  \ell_2^\star(1+e_\ell^{\star1}\theta_{i}^0 + e_\ell^{10\star}\theta_{i-1}^1 )(1-\theta_{i-1}^0),
	\end{cases}
\end{equation}
where $ r_2^\star = (x+1)r_2 -r_1$ , $\ e_r^{1\star} = \dfrac{r_2}{(x+1)r_2 -r_1} -1$, $e_r^{\star 01} = \dfrac{xr_2}{(x+1)r_2 -r_1} -1$, $\ell_2^\star = (x+1)\ell_2 -\ell_1$, $e_\ell^{\star1} = \dfrac{\ell_2}{(x+1)\ell_2 -\ell_1} -1$, $e_\ell^{10\star} = \dfrac{x\ell_2}{(x+1)\ell_2 -\ell_1} -1.$ 

The contribution of the parameter $e^{1\star}$ is the same as $d^{1\star}$ in the model with one parameter. Meanwhile, $e^{\star 01}$ only influences the particle $i$ transition rate when it is at a distance of 1 from particle $i+1$, with $\theta_i^1$ accounting for this distance. 

Using this parameterization of the jump rates, the average current for this model can be computed in a similar manner to that in~\eqref{current_model1}.\\

\noindent \underline{\textbf{Model 3:}} 
The jump rates in~\eqref{model3} can be parameterized as follows:
\begin{equation}\label{rate_md_3}
	\begin{cases}
		r_i(\boldsymbol{\eta})=\  r_i^1(\boldsymbol{\eta}) + r_i^2(\boldsymbol{\eta}) + r_i^3(\boldsymbol{\eta})\\
		\ell_i(\boldsymbol{\eta})=\  \ell_i^1(\boldsymbol{\eta}) + \ell_i^2(\boldsymbol{\eta}) + \ell_i^3(\boldsymbol{\eta})
	\end{cases}
\end{equation}
\begin{widetext}
\noindent where
	\begin{equation}
	\begin{cases}
		r_i^1(\boldsymbol{\eta}) & =\  r_3^\star(1 + f_r^{10\star}\theta_{i-1}^1 + f_r^{1\star}\theta_{i-1}^0)(1-\theta_{i}^0)
		\theta_{i}^1,\\
		r_i^2(\boldsymbol{\eta}) & =\  r_4^\star(1 + g_r^{10\star}\theta_{i-1}^1 + g_r^{1\star}\theta_{i-1}^0)(1-\theta_{i}^0)(1-\theta_{i}^1) \theta_{i}^2,\\
		r_i^3(\boldsymbol{\eta}) & =\  r_5^\star(1 + h_r^{10\star}\theta_{i-1}^1 + h_r^{1\star}\theta_{i-1}^0)(1-\theta_{i}^0)(1-\theta_{i}^1)(1-\theta_{i}^2)\\
		\ell_i^1(\boldsymbol{\eta}) & =\  \ell_3^\star(1 + f_\ell^{\star01}\theta_{i}^1 + f_\ell^{\star1}\theta_{i}^0)(1-\theta_{i-1}^0)
		\theta_{i-1}^1,\\
		\ell_i^2(\boldsymbol{\eta}) & = \  \ell_4^\star(1 + g_\ell^{\star01}\theta_{i}^1 + g_\ell^{\star1}\theta_{i}^0)(1-\theta_{i-1}^0)(1-\theta_{i-1}^1) \theta_{i-1}^2,\\
		\ell_i^3(\boldsymbol{\eta}) & =\  \ell_5^\star(1 + h_\ell^{\star01}\theta_{i}^1 + h_\ell^{\star1}\theta_{i}^0)(1-\theta_{i-1}^0)(1-\theta_{i-1}^1)(1-\theta_{i-1}^2).
	\end{cases}
\end{equation}
with
$	r_3^\star =x r_1$, $ f_r^{10\star}  = \dfrac{r_1}{r_3^\star}-1$, $f_r^{1\star} =\dfrac{xr_2}{r_3^\star}-1$, $		r_4^\star =(x+1)r_3 - r_1$, $ g_r^{10\star}  = \dfrac{r_2}{r_4^\star}-1$, $g_r^{1\star}  = \dfrac{xr_3 +r_2-r_1}{r_4^\star}-1$, 
$ r_5^\star = (x+2)r_3 -(r_2 +r_1)$, $ h_r^{10\star}  = \dfrac{r_2}{r_5^\star}-1$, $    h_r^{1\star}  = \dfrac{(x+1)r_3 -r_1}{r_5^\star}-1$, $	\ell_3^\star =x \ell_1$, $ f_\ell^{\star01}  = \dfrac{\ell_1}{\ell_3^\star}-1$,  $f_\ell^{\star1} =\dfrac{x\ell_2}{\ell_3^\star}-1$, $ \ell_4^\star =(x+1)\ell_3 - \ell_1,\ g_\ell^{\star01}  = \dfrac{\ell_2}{\ell_4^\star}-1$, $ g_\ell^{\star1}  = \dfrac{x\ell_3 +\ell_2-\ell_1}{\ell_4^\star}-1$,  $\ell_5^\star = (x+2)\ell_3 -(\ell_2 +\ell_1)$, $ 		h_\ell^{\star01}  = \dfrac{\ell_2}{\ell_5^\star}-1$, $ h_\ell^{\star1}  = \dfrac{(x+1)\ell_3 -\ell_1}{\ell_5^\star}-1$.
\end{widetext}

We will now elucidate the functioning of the rate's parameter form \eqref{rate_md_3}. The value of the rate $r_i(\boldsymbol{\eta})$ is determined by either $r_i^1(\boldsymbol{\eta})$, $r_i^2(\boldsymbol{\eta})$, or $r_i^3(\boldsymbol{\eta})$, depending on the distance between the particles $i$ and $i+1$. If the distance is 1, then $r_i(\boldsymbol{\eta})$ equals to $r_i^1(\boldsymbol{\eta})$. If the distance is 2, then $r_i(\boldsymbol{\eta})$ equals to $r_i^2(\boldsymbol{\eta})$. If the distance is greater than 2, then $r_i(\boldsymbol{\eta})$ equals to $r_i^3(\boldsymbol{\eta})$. The distances are indicated by factors $(1-\theta_i^0)\theta_i^1$, $(1-\theta_i^0)(1-\theta_i^1)\theta_i^2$, and $(1-\theta_i^0)(1-\theta_i^1)(1-\theta_i^2)$ respectively.

The contributions of parameters $f^{10\star}, f^{1\star}$; $g^{10\star}, g^{1\star}$; and $h^{10\star}, h^{1\star}$ are the same as in the models with one or two parameters.

In particular, for Model 3, one has  
\begin{equation}
	j = \rho\left(\sum_{k=1}^{3} \langle r_i^k(\boldsymbol{\eta}) \rangle_{\hat{\pi}_{gc}} - \sum_{k=1}^{3} \langle \ell_i^k(\boldsymbol{\eta}) \rangle_{\hat{\pi}_{gc}}\right),
\end{equation}
where \( r^k_i(\boldsymbol{\eta}) \) and \( \ell^k_i(\boldsymbol{\eta}) \) (see \eqref{rate_md_3}) are defined in Appendix \ref{average_current}.  

\section{Proofs}\label{proofs}

\subsection{Proof of Theorem \ref{Th1}}
The master equation governing the probability $\mathbb{P}(\boldsymbol{\eta},t)$ of the system being in configuration $\boldsymbol{\eta}$ at time $t$ is given by
\begin{align}\label{mastereq_1}
	\dfrac{d}{dt}\mathbb{P}(\boldsymbol{\eta},t) & = \sum_{i=1}^{N}\bigg[r_i(\boldsymbol{\eta}_{f}^i)\mathbb{P}(\boldsymbol{\eta}_{f}^i,t) + \ell_i(\boldsymbol{\eta}_{b}^i)\mathbb{P}(\boldsymbol{\eta}_{b}^i,t)\nonumber\\  
	& \ - (r_i(\boldsymbol{\eta}) + \ell_i(\boldsymbol{\eta}))\mathbb{P}(\boldsymbol{\eta},t)\bigg],
\end{align}
where $\boldsymbol{\eta}_{f}^i$ is the configuration preceding $\boldsymbol{\eta}$ due to a forward translocation of the $i$th particle (i.e., with coordinate $p_j^{f} = p_j-\delta_{j,i}$), and $\boldsymbol{\eta}_{b}^i$ is the configuration preceding $\boldsymbol{\eta}$ due to a backward translocation of the $i$th particle (i.e., $p_j^{b} = p_j+\delta_{j,i}$). Due to periodicity, the positions $p_i$ of the particles are taken modulo $L$, and the labels $i$ are counted modulo $N$.

If a measure $\hat{\pi}$ is the invariant distribution of the process, dividing \eqref{mastereq_1} by the stationary distribution $\hat{\pi}(\boldsymbol{\eta})$ leads to the stationary condition
\begin{equation}\label{stationcond_1}
	\sum_{i=1}^{N} \bigg[r_i(\boldsymbol{\eta}_{f}^i)\dfrac{\hat{\pi}(\boldsymbol{\eta}_{f}^i)}{\hat{\pi}(\boldsymbol{\eta})} + \ell_i(\boldsymbol{\eta}_{b}^i)\dfrac{\hat{\pi}(\boldsymbol{\eta}_{b}^i)}{\hat{\pi}(\boldsymbol{\eta})}  -(r_i(\boldsymbol{\eta}) + \ell_i(\boldsymbol{\eta}))\bigg] = 0.
\end{equation}

Expressing the measure \eqref{sm_Ising} in terms of the variables $\theta_{i}^p$ \eqref{newvar}, we obtain
\begin{equation}
	\hat{\pi}(\boldsymbol{\eta}) = \dfrac{1}{Z}\prod_{i=1}^{N}x^{\theta_i^0}.
\end{equation}
It follows that
\begin{align}
	\dfrac{\hat{\pi}(\boldsymbol{\eta}_f^{i})}{\hat{\pi}(\boldsymbol{\eta})} &=  x^{-\theta_{i-1}^0 -\theta_{i}^0+\theta_{i-1}^1},\\
	\dfrac{\hat{\pi}(\boldsymbol{\eta}_b^{i})}{\hat{\pi}(\boldsymbol{\eta})} &= x^{-\theta_{i}^0 +\theta_{i}^1-\theta_{i-1}^0  }\label{bwj}.
\end{align}

Moreover, the transition rates satisfy
\begin{align*}
	r_i(\boldsymbol{\eta}) &= r(h_i,h_{i-1}),  & \ell_i(\boldsymbol{\eta}) &= \ell(h_{i-1},h_{i}), \\
	r_i(\boldsymbol{\eta}_f^i) &= r(h_i+1,h_{i-1}-1),  & \ell_i(\boldsymbol{\eta}_b^{i}) &= \ell(h_{i-1}+1,h_{i}-1).
\end{align*}

Given the dynamics \eqref{cond_jr_ep}, it is straightforward to verify that
\begin{align}\label{Noether_Th}
	r_i(\boldsymbol{\eta}_{f}^i)\dfrac{\hat{\pi}(\boldsymbol{\eta}_{f}^i)}{\hat{\pi}(\boldsymbol{\eta})}  + \ell_i(\boldsymbol{\eta}_{b}^i)&\dfrac{\hat{\pi}(\boldsymbol{\eta}_{b}^i)}{\hat{\pi}(\boldsymbol{\eta})}   -(r_i(\boldsymbol{\eta}) + \ell_i(\boldsymbol{\eta})) \nonumber \\
	& = r_{h_{i-1}} - r_{h_i} + \ell_{h_{i}} - \ell_{h_{i-1}}.
\end{align}

By summing over $i$ from $1$ to $N$, and using the periodicity of the lattice, we obtain \eqref{stationcond_1}. Thus, $\hat{\pi}$ is indeed the invariant measure of the process, completing the proof.\\

\subsection{Proof of Theorem \ref{Th2}}

Reversibility, which corresponds to the condition of detailed balance, is given by
\begin{equation}
	r_i(\boldsymbol{\eta})\hat{\pi}(\boldsymbol{\eta}) = \ell_i(\boldsymbol{\eta}_b^i)\hat{\pi}(\boldsymbol{\eta}_b^i), \quad \text{for } i=1,\dots,N,
\end{equation}
where $\boldsymbol{\eta}_{b}^i$ represents the configuration preceding $\boldsymbol{\eta}$ due to a backward translocation of the $i$th particle, i.e., $p_j^{b} = p_j + \delta_{j,i}$.

Since the transition rates are given as $r_i(\boldsymbol{\eta}) = r(h_i, h_{i-1})$ and $\ell_i(\boldsymbol{\eta}_b^i) = \ell(h_{i-1}+1, h_i-1)$, substituting these into the detailed balance equation and applying Eq. \eqref{bwj} yields
\begin{equation}
	x^{-\theta_{i}^0 + \theta_{i}^1 - \theta_{i-1}^0} \ell(h_{i-1}+1, h_i-1) = r(h_i, h_{i-1}).
\end{equation}
Thus, 
\begin{align}
	x^{-\theta_{i}^0  + \theta_{i}^1 - \theta_{i-1}^0} & \ell(h_{i-1}+1, h_i-1)  - \ell(h_{i-1},h_i)\nonumber \\
	& = r(h_i, h_{i-1}) - \ell(h_{i-1},h_i).
\end{align}
The left-hand side of the above equation is $\ell_{h_i} - \ell_{h_{i-1}}$. Thus,
\begin{equation}\label{reversible_cond}
	\ell_{h_i} - \ell_{h_{i-1}} = r(h_i, h_{i-1}) - \ell(h_{i-1},h_i).
\end{equation}
Taking $h_i = k$ for $k=0,...,L-N$ and $h_{i-1}=0$, the above equation yields $r_k = \ell_k$. Conversely, with $r_k = \ell_k$ for $k=0,...,L-N$, Eq. \eqref{reversible_cond} holds true. This completes the proof.

\end{document}